\documentclass[twocolumn]{aastex701} 

\makeatletter
\let\frontmatter@title@above\relax
\makeatother

\usepackage{tikz}
\usepackage{amsmath}
\usepackage{siunitx}
\DeclareSIUnit\year{yr}
\DeclareSIUnit\parsecs{pc}
\usepackage{subcaption}
\usepackage{booktabs}
\usepackage{makecell}
\usepackage{adjustbox}
\usepackage{changepage}
\usepackage{soul}

\begin{document}

\title{Calibrating Galaxy Infall Times in Groups and Clusters with IllustrisTNG Simulations}

\author[0009-0008-4355-021X]{Florine Masson}
\email{florine.masson@ens.psl.eu}
\affiliation{Department of Physics and Astronomy, McMaster University, 1280 Main Street West, Hamilton, ON, L8S 3L8, Canada}
\affiliation{Department of Physics, ENS-PSL, Paris, France}

\author[0000-0003-4722-5744]{Laura C. Parker}
\email{lparker@mcmaster.ca}
\affiliation{Department of Physics and Astronomy, McMaster University, 1280 Main Street West, Hamilton, ON, L8S 3L8, Canada}

\begin{abstract}
 
The time since a galaxy first became a satellite is central to understanding how environment drives galaxy evolution, yet it cannot be measured directly. Using the TNG300 and TNG-Cluster simulations, we track satellites from $z=1$ to $z=0$ and derive a simple, redshift-dependent prescription for ${T}_{\rm{inf}}$ based on position in projected phase space and stellar mass, via symbolic regression. The resulting calibration provides continuous, observation-ready estimates of infall time across projected phase space. In projected phase space, ${T}_{\rm{inf}}$ is often well described by two components, and we provide analytic expressions for the corresponding characteristic timescales. This framework can be applied directly to spectroscopic samples to infer environmental histories in galaxy groups and clusters.

\end{abstract}


\section{Introduction} 

Environment strongly shapes galaxy evolution. In groups and clusters, galaxies interact with the hot intragroup or intracluster medium and with other satellites, altering their star formation and morphology \citep{BG2006,Wetzel_2013,Roberts_2016,oxland_2024}. As galaxies become satellites, their properties evolve from field-like to showing signatures of environmental processing on timescales that depend on their time since infall  $\rm{T}_{\rm{inf}}$ \citep{Pasquali_2019,Weinmann_2009}.

Although ${T}_{\rm{inf}}$ is not directly observable, a galaxy’s projected phase-space (PPS) position (projected radius and line-of-sight velocity relative to the host halo) correlates with ${T}_{\rm{inf}}$ \citep{Pasquali_2019,Rhee_2017, Oman_2016}. Cosmological simulations can track galaxy orbits in 3D, linking ${T}_{\rm{inf}}$ to both their true spatial and velocity coordinates as well as their observationally accessible position in PPS. These simulation-calibrated ${T}_{\rm{inf}}$ measures have been used in many observational studies \citep[e.g.,][]{Sampaio_2021,oxland_2024,Kim_2023}.

Previous studies have often divided projected phase space (PPS) into discrete zones, assigning galaxies an average infall time and associated uncertainty \citep[e.g.,][]{Rhee_2017, Pasquali_2019}. However, recent work by \cite{Dou_2025} demonstrates that the infall time distribution in most PPS regions is bimodal, implying that a single representative value per region is insufficient. In addition, earlier studies \citep{Pasquali_2019,Rhee_2017}, have highlighted the large range of ${T}_{\rm{inf}}$ at fixed PPS location.

Here, we extend previous work by using IllustrisTNG simulations to calibrate ${T}_{\rm inf}$ as a function of position in projected phase space, explicitly incorporating its dependence on stellar mass and redshift ($0 < z < 1$), and testing for any additional dependence on host halo mass. We provide a continuous expression of ${T}_{\rm inf}$ in PPS to avoid the edge effects present in the previous methods that divided PPS in discrete zones. We model the full ${T}_{\rm inf}$ distributions with analytic forms that capture these dependencies, providing a flexible tool for reconstructing galaxy environmental histories across cosmic time. This work is intentionally empirical. Projected phase space is a degenerate observable that mixes orbital histories, projection effects, and halo assembly history. Our goal is not to construct a physical model of satellite orbits, but to provide a calibrated, observation-ready estimator of infall time that captures the dominant statistical trends in simulations.

Section~\ref{sec:data} presents the IllustrisTNG simulations used in this work. In Section~\ref{sec:Tinf}, we explore the distribution of ${T}_{\rm{inf}}$ in PPS. Section~\ref{sec:formula} presents the details on how we find the best-fitting equation for infall time and the associated uncertainty and we compare to previous estimates in Section~\ref{sec:finalformula}. Section~\ref{sec:distribs} explores the full distributions of  $\rm{T}_{\rm{inf}}$ in PPS. These distributions are often well described by a two-components model and we provide an estimate for the means of the two time scales present in the distributions. We present conclusions and a summary of our work in Section~\ref{sec:ccl}.

\section{Data} \label{sec:data}

To calibrate the relationship between projected phase-space location and galaxy infall time, we use simulations that provide large statistical samples of groups and clusters spanning a broad range in halo mass.

\subsection{Simulations} \label{subsec:simulation}

IllustrisTNG is a suite of cosmological, magnetohydrodynamical simulations of galaxy formation (\cite{Nelson_2021}, \cite{Marinacci_2018}, \cite{Naiman_2018}, \cite{Nelson_2017}, \cite{Pillepich_2017}, \cite{Springel_2017}). In this work we use two runs from this suite, TNG300 and TNG-Cluster:
\begin{enumerate}
    \item {\bf TNG300:} A large-volume simulation that evolves a $(302.6\ \mathrm{Mpc})^3$ cube from $z = 127$ to $z = 0$.
    
    \item {\bf TNG-Cluster:} An extension of the IllustrisTNG project containing 352 high-resolution zoom-in re-simulations of galaxy clusters \citep{Nelson_2024}, evolved from $z = 120$ to $z = 0$.

\end{enumerate}

Both simulations adopt the same physical model, and the zoom-in targets in TNG-Cluster have the same numerical resolution as TNG300 ($m_{\mathrm{baryon}} \approx 1.1 \times 10^{7} \, M_{\odot}$, $m_{\mathrm{DM}} \approx 5.9 \times 10^{7} \, M_{\odot}$). The regions between zoom-in targets in TNG-Cluster have lower resolution.

TNG300 contains many group- and cluster-mass halos but relatively few very massive clusters ($M_{200} > 10^{14.5}\ M_\odot$). Combining TNG300 with TNG-Cluster therefore yields a large, statistically powerful sample spanning a wide halo mass range ($M_{200}\sim10^{13}$--$10^{15.5}\,M_\odot$).

The simulations provide 100 snapshots with associated group catalogs, identifying halos (groups/clusters) via a Friends-of-Friends algorithm and subhalos (galaxies) via the SUBFIND algorithm \citep{Springel_2001}. Merger trees are constructed using SubLink \citep{Rodriguez_Gomez_2015} and LHaloTree \citep{Springel_2005}.

\subsubsection{Sample} \label{subsec:sample}

For each subhalo, we extract the position, velocity, stellar mass and group number. We remove subhalos flagged as no longer physically meaningful (e.g., tidally disrupted or stripped below the resolution limit) \citep{Nelson_2021}. For each halo, we extract the central subhalo ID and virial mass ($M_{\rm h}$), assigning the position and velocity of the central galaxy to those of its host halo. In TNG-Cluster, we keep only the halos corresponding to the $352$ primary zoom-in targets, to ensure consistent resolution with TNG300. In TNG300, we keep all 3,545 groups ($10^{13}M_\odot< M_{\rm h} < 10^{14}M_\odot$) and 280 clusters ($10^{14}M_\odot< M_{\rm h} < 10^{15.2}M_\odot$) from the catalog.

Our galaxy sample includes all subhalos with stellar masses $>10^9 M_\odot$, ensuring each galaxy contains at least 90 particles (sufficient for this work where we are interested in where the subhalos are in the halo, not their detailed resolved properties) and is well-matched to observational surveys \citep{Sampaio_2021,oxland_2024}. Throughout this paper, we refer to subhalos as ``galaxies''. We restrict our analysis to galaxies associated with groups and clusters of mass $M_{\rm h} > 10^{13}M_\odot$, and that lie within three virial radii ($3R_{200}$) of the host center. This selection is based on a geometric criterion rather than on Friends-of-Friends membership, allowing the calibration to be more directly applicable to observational samples constructed with different group-finding algorithms.

In order to mimic observations, we select a line of sight (LOS) and project position data along this LOS, retaining only two spatial coordinates, and one velocity coordinate ($v_{LOS}$). We use the three principal axes ($x$, $y$, $z$) as independent LOS projections for both simulations to increase statistics. 

\begin{figure}
    \centering
    \includegraphics[width=\linewidth]{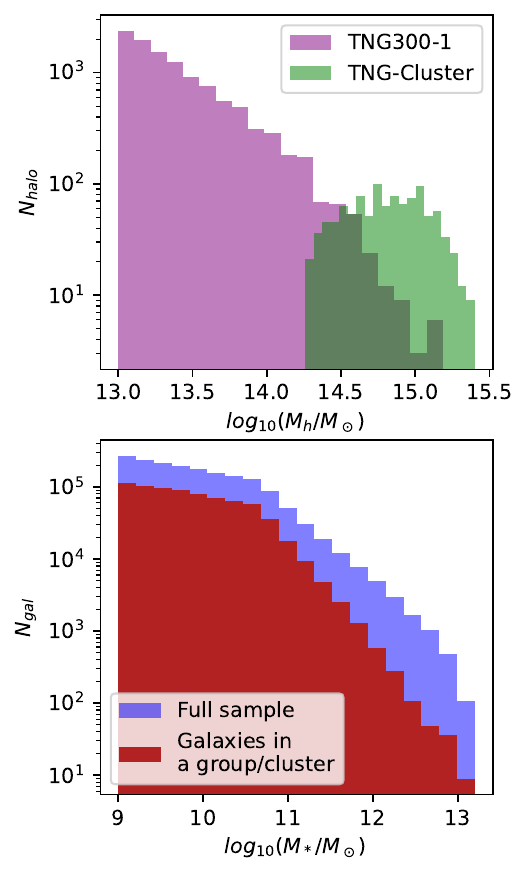}
    \caption{Distributions of halo masses (top panel) and stellar masses (bottom panel) in our sample. The red histogram in the bottom panel includes galaxies in groups and clusters with $M_{\rm h} > 10^{13}M_\odot$, while the blue histogram contains the full sample of galaxies, which includes galaxies in lower-mass halos that are not part of our selection}.

    \label{fig:distribsMhMs}
\end{figure}

Our final sample contains more than $1.5 \times 10^{6}$ galaxies (about $521{,}000$ per line of sight), of which roughly $520{,}000$ are members of groups or clusters. The other galaxies are either field galaxies or members of groups with $M_{\rm h} < 10^{13}M_\odot$. The distributions of host halo masses and galaxy stellar masses in our final sample are presented in Fig.~\ref{fig:distribsMhMs}. The top panel shows that combining the two simulations greatly increases the halo mass range we have access to.

\section{Infall time} \label{sec:Tinf}

\subsection{Definition} \label{subsec:defTinf}

The definition of infall time varies in the literature. Here, we define a galaxy as having entered a cluster or group, and thus become a satellite of this halo, when it first crosses three times the virial radius ($R_{200}$) of its current host. The corresponding infall time ($T_{\rm inf}$) is the time elapsed between this crossing and the present day, for the case of $z=0$. 

While some studies define infall time as the moment a galaxy first crosses $R_{200}$, we instead adopt $3R_{200}$ for two reasons. First, environmental effects (e.g., ram-pressure stripping, tidal forces) are known to influence galaxies beyond $R_{200}$ \citep{Balogh_2000,Oman_2020,Haines_2015}. Second, because our analysis examines projected phase space (PPS) out to $3R_{200}$, adopting $R_{200}$ would introduce an artificial discontinuity in the $T_{\rm inf}$ distribution at $R_{200}$ in PPS, complicating model fitting. This choice is discussed further in Section~\ref{subsec:PPS}.

\subsection{Galaxy tracking} \label{subsec:compMethGetTinf}

We measure $T_{\rm inf}$ using SubLink merger trees from the IllustrisTNG database, which provide galaxy positions and the time evolution of host halo radii. For each galaxy, we compute its 3D distance from the central galaxy. If the galaxy lies within $3R_{200}$ of its host at the reference redshift (e.g., $z=0$), we then trace its normalized group-centric distance, $R_{3D}(z)/R_{200}(z)$, backward through all earlier snapshots. The earliest snapshot at which this ratio first falls below three defines the infall redshift, and the corresponding lookback time from the reference redshift gives the galaxy’s $T_{\rm inf}$.

\subsection{Projected phase space (PPS)} \label{subsec:PPS}
We construct projected phase-space (PPS) diagrams including all satellite galaxies where PPS coordinates are defined as
\[
r = \frac{R_{\rm proj}}{R_{200}}, \quad v = \left| \frac{v_{\rm LOS}}{\sigma} \right|,
\]
where $\sigma$ is the halo velocity dispersion, computed following \citet{Yang_2007}:
\begin{equation}
    \sigma = \SI{397.9}{\kilo\meter\per\second} \left( \frac{M_h}{10^{14}h^{-1}M_\odot} \right)^{0.3214}.
\end{equation}

\begin{figure}
    \centering
    \includegraphics[width=\linewidth]{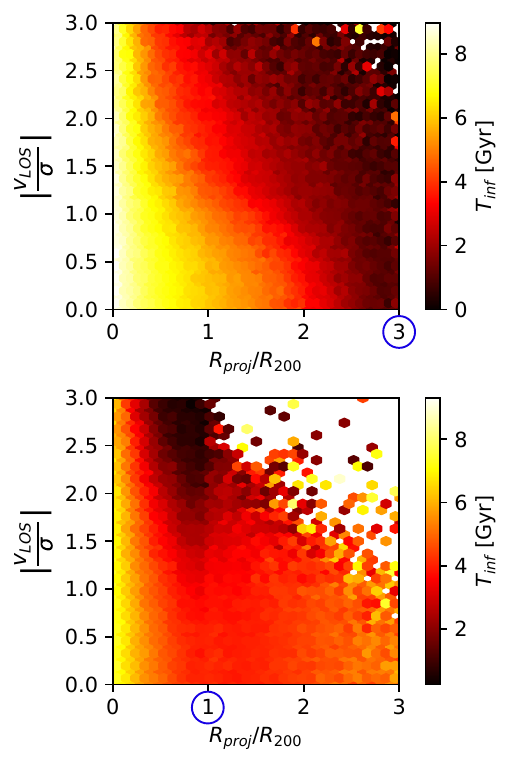}
    \caption{Projected phase-space (PPS) diagrams colour-coded by infall time, $T_{\rm inf}$, for all galaxies in the sample. In the top panel, $T_{\rm inf}$ is defined as the time when a galaxy first crosses $3R_{200}$; in the bottom panel, it is defined at $R_{200}$. The corresponding radii are marked by blue circles.}
    \label{fig:PPSsimple}
\end{figure}

Fig.~\ref{fig:PPSsimple} compares the $T_{\rm inf}$ distributions obtained when defining infall at $3R_{200}$ (top) versus $R_{200}$ (bottom). Using $R_{200}$ introduces a clear discontinuity, at $r=1$ in PPS, making interpretation difficult. The $3R_{200}$ definition avoids this issue and includes galaxies that are already environmentally affected at large radii. In both panels, galaxies are colour-coded by the mean $T_{\rm inf}$ within each PPS bin.

\subsection{Interlopers}

Due to projection effects, the PPS region shown in Fig.~\ref{fig:PPSsimple} is contaminated by galaxies located beyond $3R_{200}$ from the halo center in three-dimensional space. We classify as \textit{interlopers} those galaxies whose true 3D distance from the halo centre lies between $3R_{200}$ and $10R_{200}$ (following \citealt{Dou_2025}) but whose projected position falls within $3R_{200}$. Such galaxies can mimic genuine satellites in PPS despite never having experienced comparable environmental conditions, potentially biasing inferred infall times and quenching trends. All interlopers were removed prior to constructing the distributions shown in Fig.~\ref{fig:PPSsimple}.

\begin{figure}
    \centering
    \includegraphics[width=\linewidth]{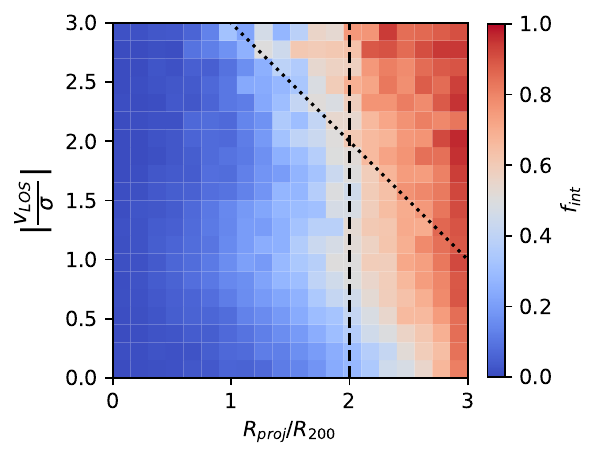}
    \caption{Fraction of interloper galaxies in projected phase space (PPS) at $z=0$. Interlopers are defined as galaxies located between $3R_{200}$ and $10R_{200}$ in 3D space but projected within $3R_{200}$. The contamination fraction peaks at large projected radii ($r \gtrsim 2$) and high relative velocities, above and to the right of the black dashed and dotted lines. The dashed line marks $r=2$, and the dotted line marks $r+v=4$, above which there are too few galaxies in our sample for robust statistical analysis, as explained in Sec.~\ref{sec:distribs}.}
    \label{fig:PPSinterlop}
\end{figure}

Fig.~\ref{fig:PPSinterlop} shows how the interloper fraction varies across PPS. As found in previous studies \citep[e.g.,][]{Rhee_2017, Dou_2025}, contamination is highest at large radii ($r \gtrsim 2$) and high relative velocities, where interlopers can account for more than 60\% of apparent satellites. In the central regions ($r \lesssim 1$), the fraction falls below 10\%. We exclude interlopers (i.e., galaxies with true 3D distance $>3R_{200}$ but projected within $3R_{200}$) from our analysis, since we have no reliable estimate of their infall times. Observationally, contamination increases sharply at $r>2$; we therefore recommend restricting analyses to galaxies within $r<2$.

\section{A formula for infall time} \label{sec:formula}

Our goal is to derive a simple, observationally applicable formula for $T_{\rm inf}$ as a function of projected phase-space position, stellar mass, and redshift. Halo mass was also tested but found to have negligible predictive power.

\subsection{$z=0$ calibration} \label{subsec:form-z0}

Infall time varies systematically with position in PPS, but each location in PPS exhibits substantial intrinsic scatter in $T_{\rm inf}$. For practical use with spectroscopic samples, we seek a continuous mapping that captures the dominant dependencies with as few parameters as possible. We begin at $z=0$ to identify an analytic relation between $T_{\rm inf}$ and a small set of predictors accessible to observers. Specifically, we consider the predictors $(r, v, \log_{10} M_\ast, \log_{10} M_{\rm h})$, where $M_\ast$ is stellar mass and $M_{\rm h}$ is host halo mass. Although previous studies (e.g., \cite{Pasquali_2019}, \cite{Rhee_2017}) showed that the dependencies of infall time on stellar and host halo masses are weak, we included these parameters to test whether even small trends could improve the predictive performance of the calibration.

We model
\[
T_{\mathrm{inf}} = f\!\left(r,\, v,\, \log_{10} M_\ast,\, \log_{10} M_{\mathrm{h}}\right),
\]
and determine $f$ in two stages. First, we coarsely grid the four-dimensional space and use symbolic regression to identify a compact functional form; second, we refit the coefficients of the best fitting function using all individual galaxies (with interlopers removed).

We use the \texttt{PySR} symbolic regression package \citep{Cranmer_2023_PySR} to search for functional forms relating $T_{\rm inf}$ to the four parameters. Symbolic regression provides an interpretable, data-driven way to uncover analytic relationships between projected phase-space coordinates and infall time, allowing us to capture the principal trends without imposing a predefined model. To keep the computation tractable, we sample the 4D space $(r, v, \log_{10}(M_\ast/M_\odot), \log_{10}(M_{\rm h}/M_\odot))$ on regular grids, replacing all galaxies in each cell with a single point defined by the mean parameter values and mean infall time of that cell. Grid sizes range from $20\times20\times8\times8$ down to $8\times8\times6\times1$ cells (dimensions listed in the order $r$, $v$, $\log M_\ast$, $\log M_{\rm h}$). In this analysis, each cell is weighted by $\log_{10}(N_{\rm gal})$, where $N_{\rm gal}$ is the number of galaxies in the cell, so that sparsely populated regions have less impact on the fit. Results are consistent without weighting, albeit with slightly larger scatter.

To control model complexity and avoid overfitting, we restricted the symbolic–regression search space to the monotonic operators $(+, -, \times, /, \log, \exp, \sqrt{\ })$ and forbid highly non-linear expressions (e.g., $ r^v$). We repeated the search across multiple grid resolutions. Across these trials, only a small minority of candidate functional forms retained a dependence on halo mass. As described in the next Section, comparison among all candidate models showed that those including $M_{\rm h}$ do not provide a better fit to the data than $M_{\rm h}$–independent alternatives. A complementary correlation analysis of the full, unbinned sample likewise indicated that $M_{\rm h}$ adds negligible explanatory power once $(r, v, M_\ast)$ are included. Consequently, we removed $M_{\rm h}$ from the calibration: subsequent grid searches were performed without a halo–mass dimension, and all remaining fits and results in this paper are based on $(r, v, M_\ast)$ only. 

\subsubsection{Best-fit functional form}\label{subsubsec:TinfFunction}

We rank candidate expressions by their root-mean-square error (RMSE, {$\sigma_{T_{\mathrm{inf}}}$) and select the lowest-RMSE form at $z=0$ returned by \texttt{PySR}:

\begin{equation} \label{eq:Tinf_neg}
    \frac{T_{\mathrm{inf}}}{\SI{1}{\giga\year}} 
    = -v + \frac{A - \log_{10}\!\left(\frac{M_*}{10^{10} M_\odot}\right)}{B + r^{C}}.
\end{equation}

We clip nonphysical negative predictions to zero: 

\begin{equation} \label{eq:Tinf}
    \frac{T_{\mathrm{inf}}}{\SI{1}{\giga\year}} 
    = \max\!\left(0,\; -v + \frac{A - \log_{10}\!\left(\frac{M_*}{10^{10} M_\odot}\right)}{B + r^{C}}\right).
\end{equation}

After identifying the functional form using the gridded data, the coefficients were fitted using the full (interloper-free) galaxy sample, yielding
\[
A = 14.91\pm0.05,\quad B = 1.615\pm0.007,\quad C = 1.347\pm0.006 .
\]

We note that this functional form has trends with $r$ and $v$ consistent with qualitative expectations: infall time increases toward smaller projected radii and smaller velocities, corresponding to galaxies that are deeper within their host potential. These trends are recovered in Eq.~(\ref{eq:Tinf}). We emphasize that this functional form should be interpreted as a calibrated estimator rather than a physical model.

\subsubsection{Performance at \(z=0\)} \label{subsubsec:rmse_z0}

\begin{figure}
    \centering
    \includegraphics[width=\linewidth]{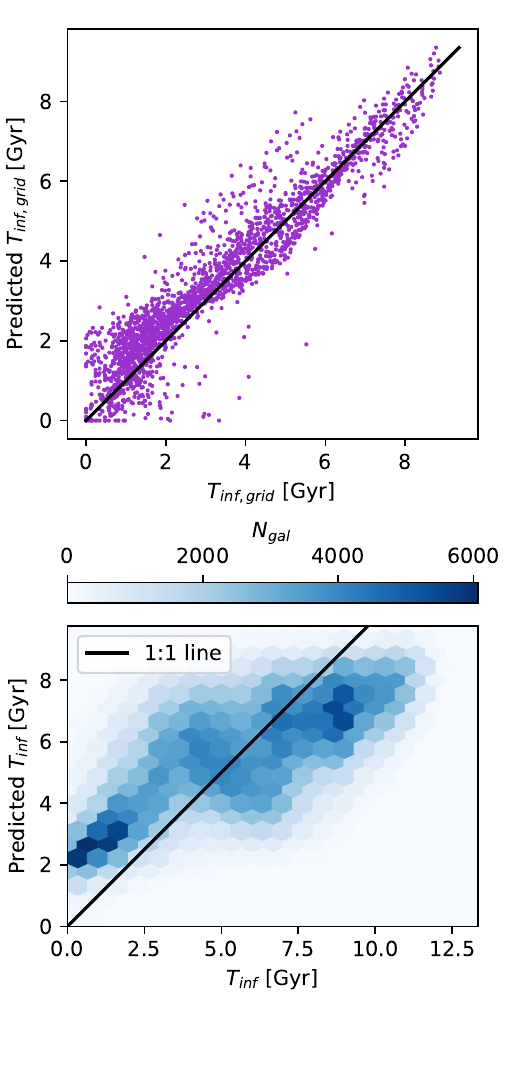}

\caption{Comparison between the true infall time measured directly from the simulation and the value predicted by Eq.~(\ref{eq:Tinf}) at $z=0$. In the top panel, each dot corresponds to the mean value of a grid cell (grid dimensions of 20x20x6 in $(r, v, M_\ast)$). The 2D histogram in the bottom panel shows the distribution for individual galaxies. In both cases, the coefficients $A$, $B$, and $C$ are those fitted using the full (interloper-free) galaxy sample. The black lines indicate the one-to-one relation.}
    \label{fig:compMeanAndInd_RealComp}
\end{figure}

Fig.~\ref{fig:compMeanAndInd_RealComp} compares the predicted $T_{\rm inf}$ from Eq.~(\ref{eq:Tinf}) with the true infall times measured directly from the simulation at $z=0$. In the top panel (purple points), each point shows the mean $T_{\rm inf}$ within a grid cell (grid resolution $20 \times 20 \times 6$ in $(r, v, M_\ast)$), while the predicted values are computed using coefficients refit on individual galaxies. The bottom panel presents the same comparison for individual galaxies. In both cases, the model captures the overall trend but compresses the dynamic range: short $T_{\rm inf}$ are biased high and long $T_{\rm inf}$ are biased low. This compression is particularly evident in the two-dimensional histogram of individual galaxies (bottom).

For individual-galaxy predictions (bottom panel), with the A, B and C values given above, the overall RMSE is \SI{2.31}{\giga\year} across the PPS. As shown in Fig.~\ref{fig:gridRMSE}, absolute errors are smallest in the upper-right region of PPS, where typical $T_{\rm inf}$ values approach zero; however, the fractional error can be large (RMSE$/T_{\rm inf}\gtrsim 1$, see Appendix~\ref{app:rmse(PPS)(z)}). Conversely, absolute errors increase toward the lower-left region, but they represent a smaller fraction of the true value because $T_{\rm inf}$ can reach up to \SI{13.3}{\giga\year}.

\begin{figure}
    \centering
    \includegraphics[width=\linewidth]{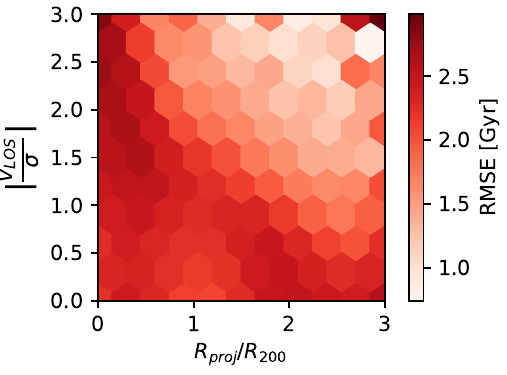}
    \caption{RMSE in the prediction of infall time as a function of position in PPS for the $z=0$ sample.}
    \label{fig:gridRMSE}
\end{figure}

\subsection{Redshift evolution of the calibration} \label{subsubsec:from-severalz}

At higher redshifts, less cosmic time has elapsed and clusters are still assembling, so the infall-time calibration may evolve with redshift. We first test whether the same functional form (Eq.~(\ref{eq:Tinf})) remains valid at earlier epochs by examining snapshots at \(z \in \{0, 0.2, 0.3, 0.5, 0.7, 1\}\). To test this, we refit the coefficients \(A(z)\), \(B(z)\), and \(C(z)\) in Eq.~(\ref{eq:Tinf}) at each redshift using {\fontfamily{cmtt}\selectfont curve\_fit}, and compute alternative candidate functional forms at each redshift following the method described in Section~\ref{subsec:form-z0}. We then compare the RMSE of infall times predicted by Eq.~(\ref{eq:Tinf}) with the refitted coefficients to those of the alternative fits. We find that Eq.~(\ref{eq:Tinf}) continues to provide a good description up to \(z = 1\), with RMSEs similar to or lower than the alternatives.

Quadratic functions of \(z\) describe the evolution of \(A\) and \(B\), while a linear function suffices for \(C\)  (Fig.~\ref{fig:ABC}). Incorporating these redshift-dependent coefficients reduces the RMSE compared to using a fixed \(z = 0\) calibration at all redshifts.

\begin{figure*}[t]
    \centering
    \includegraphics[width=\textwidth]{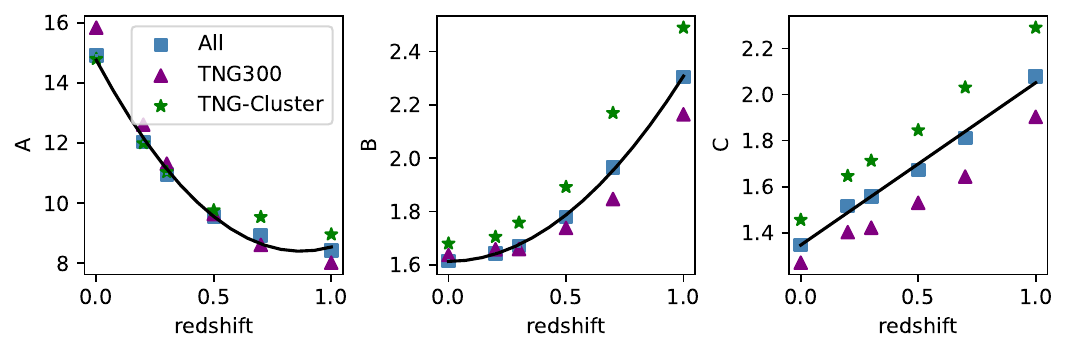}
     \caption{Coefficients $A$, $B$, and $C$ as functions of redshift. The black lines show the best-fit relations for the combined TNG-Cluster and TNG300 data (blue squares). Quadratic polynomials are used for $A$ and $B$, and a linear fit for $C$. Error bars are present but smaller than the marker size.}
    \label{fig:ABC}
\end{figure*}

The redshift dependence is:

\begin{equation} \label{eq:A}
    A(z) = 8.384z^2 - 14.603z + 14.758,
\end{equation}

\begin{equation} \label{eq:B}
    B(z) = 0.694z^2 - 0.001z + 1.613,
\end{equation}

\begin{equation} \label{eq:C}
    C(z) = 0.704z + 1.347.
\end{equation}

Fig.~\ref{fig:ABC} shows the fitted coefficients for TNG300 and TNG-Cluster separately, and for the combined dataset. Although the absolute values differ slightly between simulations, the redshift trends are consistent, and the RMSE computed using the combined or single-simulation coefficients differs by less than $\SI{0.06}{\giga\year}$ at each redshift. The increasing difference between the coefficients fitted to TNG300 and TNG-Cluster at high redshift, especially for coefficient $B$ (Fig.~\ref{fig:ABC}), could indicate a modest increase in sensitivity to halo mass at earlier epochs. However, because these differences have a negligible impact on the RMSE, we do not pursue this possibility further here.

Fig.~\ref{fig:rmse(z)} shows the evolution of the mean RMSE across the PPS with redshift. The RMSE decreases steadily with increasing redshift, reflecting the shorter cosmic times available for infall at earlier epochs. Averaged over the entire PPS region, uncertainties are ~\SI{2.5}{\giga\year} at $z=0$ and decline to ~\SI{1}{\giga\year} at $z=1$. The RMSE evolution with redshift is well described by:
\begin{equation}
    \frac{\rm{RMSE}(z)}{\SI{1}{\giga\year}} = 1.328z^2 - 2.737z + 2.281 .
\end{equation}

Variations in RMSE across PPS at each redshift are shown in Appendix~\ref{app:rmse(PPS)(z)}, Fig.~\ref{fig:rmse(z)PPS}.

\begin{figure}
    \centering
    \includegraphics[width=\linewidth]{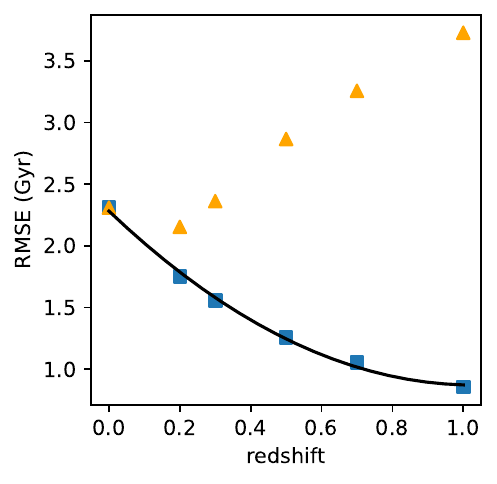}
   \caption{Mean RMSE of the predicted infall times, averaged over all galaxies in the projected phase space, as a function of redshift. The blue squares are the RMSEs computed using the A, B and C coefficients computed with Eqs.\ref{eq:A}-\ref{eq:C}. The orange triangles are the RMSEs computed with A(z=0), B(z=0), C(z=0).}
    \label{fig:rmse(z)}
\end{figure}

If the $z=0$ coefficients are applied at higher redshift without refitting, the RMSE systematically increases with redshift, as shown by the orange triangles in Fig.~\ref{fig:rmse(z)}. These results underscore the importance of incorporating explicit redshift dependence into the $T_{\rm inf}$ calibration.

\section{Final Calibration and Comparison to Previous Work} \label{sec:finalformula}

A primary goal of this work is to provide an updated calibration of infall time that includes an explicit dependence on stellar mass and extends to higher redshift. The final calibration,
\begin{equation} \label{eq:Tinf(z)}
    \frac{T_{\mathrm{inf}}}{\SI{1}{\giga\year}} 
    = \max\!\left(0,\; -v + \frac{A(z) - \log_{10}\!\left(\frac{M_*}{10^{10} M_\odot}\right)}{B(z) + r^{C(z)}}\right),
\end{equation} 
uses the redshift-dependent coefficients \(A(z)\), \(B(z)\), and \(C(z)\) derived in Section~\ref{subsubsec:from-severalz}. This relation can be applied directly to observed samples for which projected radius, line-of-sight velocity, and stellar mass are known.

If stellar mass measurements are unavailable, or when a simplified prescription is preferred given the weak dependence on $M_\ast$, we provide a simplified version of the calibration in which the $M_\ast$-dependent term is replaced by the mean stellar mass of the sample in Eq.~(\ref{eq:Tinf(z)}):
\begin{equation}\label{eq:Tinf(z)_meanMs}
    \frac{T_{\mathrm{inf}}}{\SI{1}{\giga\year}} 
    = \max\!\left(0,\; -v + \frac{A(z) - 10.43}{B(z) + r^{C(z)}}\right).
\end{equation} 

The RMSEs obtained with this version are similar, but slightly higher than those derived from Eq.~(\ref{eq:Tinf(z)}). They are provided in Table \ref{tab:rmseMsFixed} in Appendix \ref{app:MsDependence}.

\subsection{Comparison to previous calibrations at $z=0$}

\citet[hereafter P19]{Pasquali_2019} divided projected phase space into eight zones out to $R_{200}$ and calibrated mean infall times within each region using simulations. Their definition of infall time corresponds to a galaxy’s first crossing of $R_{200}$, whereas our calibration adopts $3R_{200}$ as the entry radius. Because of this difference, the absolute $T_{\rm inf}$ values are not directly comparable, but the relative trends can be examined.

Table~\ref{tab:comparison_pasquali} summarizes the mean $T_{\rm inf}$ predicted by our model within the P19 zones (excluding galaxies with $r>1$ for consistency). The progression of mean $T_{\rm inf}$ across zones closely matches that reported by P19. The mean difference of 2.3 to 3.3 Gyr between the two definitions corresponds roughly to the time required for a galaxy to travel from $3R_{200}$ to $R_{200}$ on first infall.

\begin{deluxetable*}{lccccccccccc}
\tablecaption{Mean infall times in the Pasquali et al. (2019; P19) zones and RMSE of our calibration as a function of redshift.\label{tab:comparison_pasquali}}
\tablehead{
\colhead{Zone\tablenotemark{a}} &
\colhead{$\overline{T}_{\mathrm{inf,eq},z=0}$\tablenotemark{b}} &
\colhead{$\overline{T}_{\mathrm{inf,P19}}$\tablenotemark{c}} &
\colhead{$\Delta T$\tablenotemark{d}} &
\multicolumn{7}{c}{$\sigma_{T_{\mathrm{inf}}}$ [Gyr]\tablenotemark{e}} \\
\colhead{} &
\colhead{[Gyr]} & \colhead{[Gyr]} & \colhead{[Gyr]} &
\colhead{P19} & \colhead{$z=0.0$} & \colhead{$z=0.2$} &
\colhead{$z=0.3$} & \colhead{$z=0.5$} & \colhead{$z=0.7$} & \colhead{$z=1.0$}
}
\startdata
1 & 8.41 & 5.42 & 2.99 & 2.51 & 2.30 & 1.82 & 1.66 & 1.35 & 1.12 & 0.95 \\
2 & 7.67 & 5.18 & 2.49 & 2.60 & 2.43 & 1.93 & 1.76 & 1.46 & 1.23 & 1.01 \\
3 & 6.99 & 4.50 & 2.49 & 2.57 & 2.44 & 1.92 & 1.74 & 1.44 & 1.24 & 1.04 \\
4 & 6.20 & 3.89 & 2.31 & 2.34 & 2.37 & 1.83 & 1.63 & 1.33 & 1.14 & 0.95 \\
5 & 5.70 & 3.36 & 2.34 & 2.36 & 2.45 & 1.87 & 1.67 & 1.35 & 1.13 & 0.88 \\
6 & 5.38 & 2.77 & 2.61 & 2.29 & 2.47 & 1.90 & 1.68 & 1.33 & 1.13 & 0.88 \\
7 & 5.12 & 2.24 & 2.88 & 1.97 & 2.40 & 1.82 & 1.61 & 1.29 & 1.07 & 0.84 \\
8 & 4.69 & 1.42 & 3.27 & 1.49 & 2.12 & 1.60 & 1.40 & 1.18 & 1.09 & 0.95 \\
\enddata
\tablenotetext{a}{Zone number from P19.}
\tablenotetext{b}{Mean infall time computed with Eq.~\ref{eq:Tinf} at $z=0$.}
\tablenotetext{c}{Mean infall time from P19.}
\tablenotetext{d}{Difference between our mean infall time and that of P19.}
\tablenotetext{e}{Root-mean-square error (RMSE) of predicted infall times from P19 and from our calibration at each redshift.}
\end{deluxetable*}

We compare the root-mean-square errors (RMSEs) in $T_{\rm inf}$ between our calibration and that of P19 at $z=0$ by evaluating Eq.~(\ref{eq:Tinf(z)}) within the zones defined by P19. The $z=0$ RMSEs from our model and those reported by \citet{Pasquali_2019} (fifth column) are listed in Table~\ref{tab:comparison_pasquali}. Our RMSEs are similar to P19’s in the inner zones and somewhat larger in zones~7–8. This difference may reflect the weighting adopted in our calibration, which places greater statistical emphasis on the more densely populated inner regions of PPS. Alternative weighting schemes were explored, but none produced a lower global RMSE. Because our calibration defines infall at $3R_{200}$ rather than $R_{200}$, the corresponding $T_{\rm inf}$ values are systematically higher with our method. Although our fit includes stellar mass, the overall RMSE values remain comparable to those of P19. The remaining scatter highlights the intrinsic difficulty of reducing the complex infall-time distributions to a single representative value per zone. For completeness, Table~\ref{tab:comparison_pasquali} also reports our RMSEs at higher redshifts, which are not available in P19.

Although our calibration does not yield lower RMSE than P19 at $z=0$, its continuous functional form eliminates edge effects associated with discrete zone boundaries. Our framework extends naturally to higher redshift, making it directly applicable to other observational samples.

\section{Infall Time Distributions at \lowercase{$z$}=0}\label{sec:distribs}

While the single-value infall time model from Section~\ref{sec:formula} achieves lower relative errors than zone–based approaches \citep{Pasquali_2019}, its prediction uncertainty remains a substantial fraction of the infall time itself.  One reason for this limitation is that the infall time distribution in projected phase space (PPS) is often better described as a sum of several components rather than a single one, as noted by \citet{Dou_2025}. They showed that dividing the PPS into a number of bins reveals distinct populations of early and recent infallers. A single functional form cannot fully capture this structure, which likely contributes to the high RMSE in our predictions.  

In this section, we examine the ${T_{\rm inf}}$ distribution at each PPS location at $z=0$ and model it as the sum of two Gaussian components representing early- and late-infall populations. Rather than adopting fixed “zone” boundaries, each Gaussian component is parameterized as a continuous function of $(r, v, \log_{10}(M_*/M_\odot))$.

\subsection{Illustration of the two-component fitting}\label{subsec:illustDistrib}

As in Section~\ref{subsec:form-z0}, we divide the three-dimensional space $(r, v, \log_{10}(M_*/M_\odot))$ into regular cells, using the same range of grid sizes as before. In each cell, we compute the distribution of infall times and fit it as the sum of two Gaussian components using the {\fontfamily{cmtt}\selectfont GaussianMixture} implementation from the {\fontfamily{cmtt}\selectfont scikit-learn} library \citep{scikit-learn}. An example for a $6\times6\times1$ grid is shown in Fig.~\ref{fig:gridHist}. The values of the Gaussian parameters used to plot the black curves of Fig.~\ref{fig:gridHist} are listed in Appendix~\ref{app:gaussianParameters}, Table~\ref{tab:grid66gaussians}.

\begin{figure*}[t]
    \centering
    \includegraphics{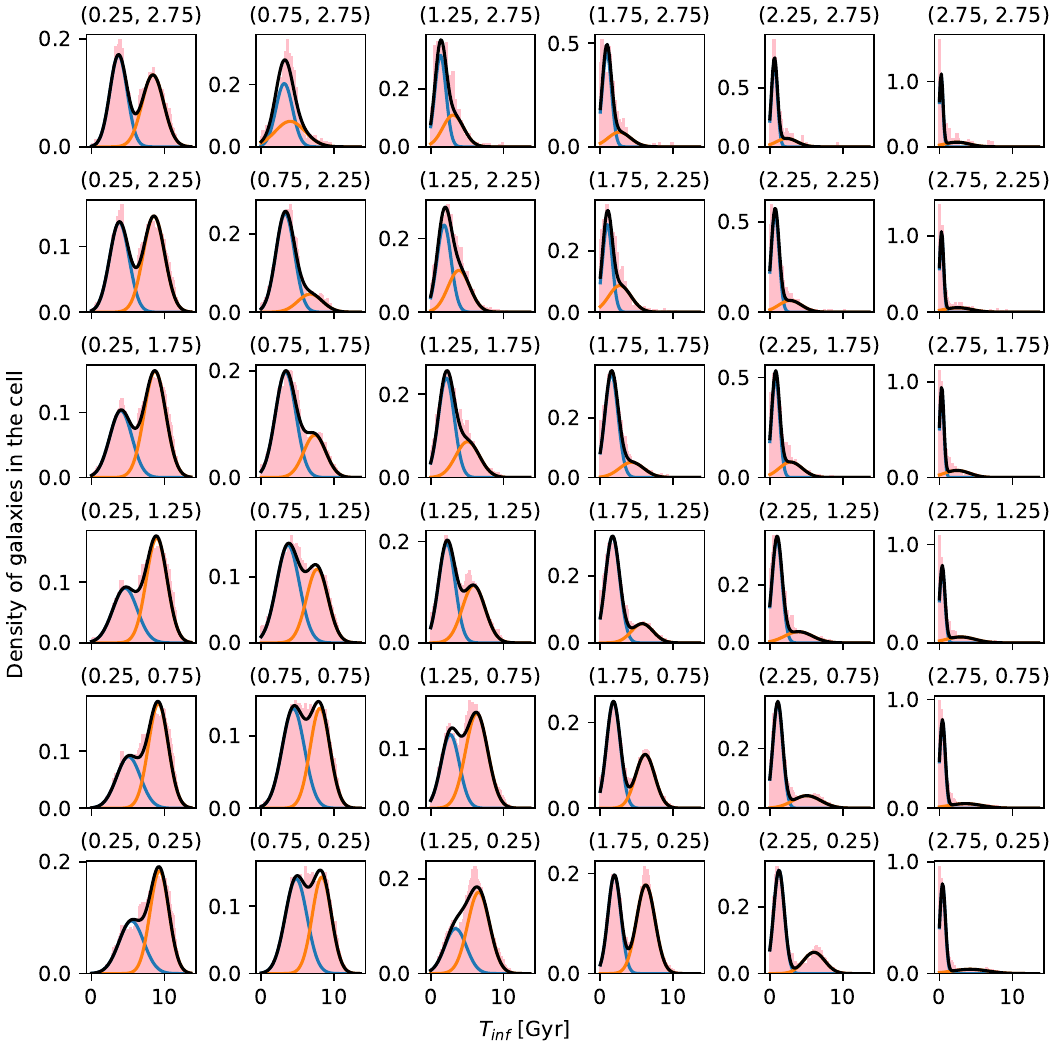}
    \caption{Infall-time distributions for cells in projected phase space (PPS) using a $6\times6\times1$ grid (dimensions of  $r$, $v$, $\log M_\ast$), combining all stellar and halo masses. Panel positions correspond to each cell's $(r, v)$ location in PPS. The blue and orange curves show the two Gaussian components, and the black curve is their sum. Panel titles list the cell centers $(r,v)$, corresponding to $(\tfrac{r_{\rm max}+r_{\rm min}}{2},\,\tfrac{v_{\rm max}+v_{\rm min}}{2})$.}
    \label{fig:gridHist}
\end{figure*}

In most PPS regions, the distributions are well described by a sum of two components. In cells with $v \gtrsim 1.5$, a single Gaussian component is often sufficient. Cells at very large radii and velocities ($r+v > 4$) contain relatively few galaxies and should be interpreted with caution.

\subsection{Functional forms for the two means} \label{subsec:fitTwoMeans}

We use {\fontfamily{cmtt}\selectfont PySR} \citep{Cranmer_2023_PySR} to derive symbolic expressions for the two Gaussian means, $T_{\mathrm{mean},1}$ and $T_{\mathrm{mean},2}$, as functions of $(r, v, M_*)$. Note $T_{\mathrm{mean},1}$ is always the component with the smaller ${T_{\rm inf}}$. The best-fit expressions are:
 
\begin{equation} \label{eq:m1}
    \frac{T_{\rm mean,1}}{\SI{1}{\giga\year}} = \frac{(3.981-r)^{2.272}}{v+\sqrt{log_{10}(M_*/M_\odot)}},
\end{equation}

\begin{equation}\label{eq:m2}
    \frac{T_{\rm mean,2}}{\SI{1}{\giga\year}} = - 4.723\sqrt{r} - v - log_{10}(M_*/M_\odot) + 22.283.
\end{equation}

The average RMSE values across PPS, computed using a $20\times20\times6$ grid, are
$\rm RMSE_1 = \SI{0.56}{\giga\year}$ and $\rm RMSE_2 = \SI{1.20}{\giga\year}$.
We use gridded cells ($20\times20\times6$) here to allow measurement of full distributions within each region. Fig.~\ref{fig:meansRealVSComputed} shows that the measured Gaussian means and those predicted by Eqs.~\ref{eq:m1}–\ref{eq:m2} agree well.

The functional forms of Eqs.~\ref{eq:m1}–\ref{eq:m2} differ substantially from that of Eq.~(\ref{eq:Tinf(z)}). They likely reflect a combination of projection effects in projected phase space (PPS) and variations in accretion histories (e.g., early versus late infall populations), but we do not attempt to disentangle these contributions here. We emphasize that these relations should not be interpreted as a physical model of accretion. They provide an empirical decomposition of the infall-time distribution, illustrating that a single-valued estimator (Eq.~(\ref{eq:Tinf(z)})) cannot fully represent what is intrinsically a multi-valued distribution.

\begin{figure}
    \centering
\includegraphics{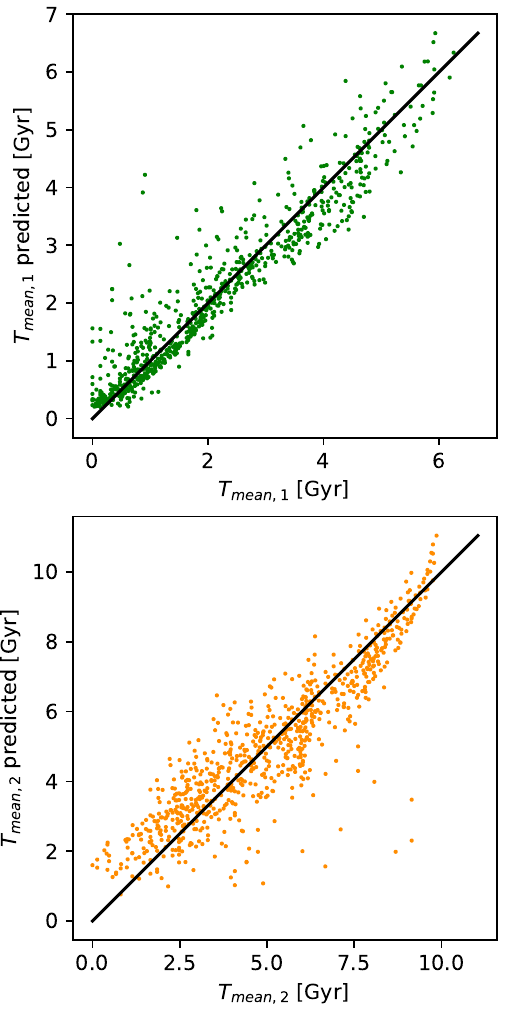}
\caption{Comparison between the measured means of the two Gaussian components and the predictions from Eqs.~\ref{eq:m1} and~\ref{eq:m2}. \textit{Top:} lower mean ($T_{\mathrm{mean},1}$); \textit{bottom:} higher mean ($T_{\mathrm{mean},2}$). The black line indicates the $1{:}1$ relation.}
    \label{fig:meansRealVSComputed}
\end{figure}

\subsection{Other Gaussian fitting parameters}
In principle, the full infall-time distribution at each PPS location could be predicted by fitting functional forms not only for the means but also for the standard deviations and relative weights of the two Gaussian components. Although the distributions appear bimodal and are well described by double Gaussians, our attempts to model the variation of these additional parameters across PPS were unsuccessful: all resulting models provided poor fits to the data, with RMSEs comparable to or larger than the intrinsic scatter of the data. This suggests that the variables considered here, $(r, v, M_*)$, are insufficient to fully capture the shape of the underlying distributions.

Fig.~\ref{fig:w1grid} shows the weight of the Gaussian with the smaller mean $\rm{T}_{\rm{inf}}$ in PPS. A weight near unity indicates that the distribution is dominated by the first component. Note that regions with $r > 2$ or $r+v > 4$ are sparsely populated and more affected by interlopers in observational studies.

\begin{figure}
    \centering
    \includegraphics[width=\linewidth]{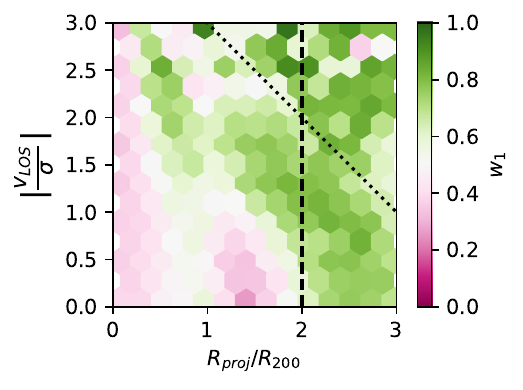}
    \caption{Weight of the Gaussian with the smallest mean, $T_{\mathrm{mean},1}$, in PPS. The black dotted and dashed lines mark $r+v=4$ and $r>2$ respectively; points above or to the right of these lines are less reliable due to low statistics and higher interloper contamination in observations}   
    \label{fig:w1grid}
\end{figure}

\subsection{Comparison of the two approaches}

A two-Gaussian description in PPS successfully captures distinct early- and late-infall populations and reduces the RMSE by roughly a factor of two, provided the correct component is identified. However, the variables used here do not reliably predict which component an individual galaxy is in. In short, while the single-value calibration offers a practical, redshift-dependent tool for estimating infall times, explicitly incorporating the bimodality has the potential to further reduce scatter in future work. The next step is to identify additional physical parameters that can distinguish between the two populations and enable more precise reconstruction of environmental histories.

\section{Summary and Conclusions} \label{sec:ccl}
We combined the TNG-Cluster and TNG300 simulations to construct a large sample of galaxy groups and clusters. By tracing member galaxies through time, we measured their infall times into their present-day halos and used symbolic regression to derive an analytic expression for infall time as a function of observable, environment-dependent parameters: projected radius ($R_{\mathrm{proj}}$), line-of-sight velocity ($v_{\mathrm{LOS}}$), and stellar mass ($M_*$). We quantified the model's performance across projected phase space (PPS) and found that halo mass does not significantly influence the relation. The resulting functional form was calibrated over the redshift range $0 \leq z \leq 1$.

We then examined the full infall-time distribution at each PPS location, modeling it as the sum of two Gaussian components. This revealed clear bimodality in many regions of PPS, consistent with previous studies, and we derived analytic expressions for the means of the two Gaussian components. However, the variables used here are insufficient to reliably determine which of the two characteristic infall times applies to a given galaxy.

This work advances previous studies by:
\begin{enumerate}
    \item Providing a calibrated, redshift-dependent formula for infall time as a function of ($R_{\mathrm{proj}}$, $v_{\mathrm{LOS}}$, $M_*$) for $0 \leq z \leq 1$ (Section~\ref{subsubsec:TinfFunction}--\ref{subsubsec:from-severalz});
    \item Explicitly incorporating stellar mass and redshift into the functional form for $T_{\mathrm{inf}}$;
    \item Showing that the infall-time distribution in many regions of PPS is well described by two components, and providing analytic expressions for the corresponding mean values (Section~\ref{sec:distribs}).
\end{enumerate}

These results provide a flexible calibration of infall time that can be easily applied to observational surveys across a range of redshifts, while highlighting the complexity of the underlying distributions that must be considered in future work.

\begin{acknowledgments}

The authors thank J.Yeung for his introduction to the IllustrisTNG database, as well as M.Oxland, M.Bravo, L.Foster and D.Lazarus for their advice and discussions.

This work was made possible thanks to the IllustrisTNG database (\cite{Nelson_2021}). 
The analysis was made possible in great part by publicly available software packages, including Numpy (\cite{Harris_2020}), Matplotlib (\cite{matplotlib}), PySR (\cite{Cranmer_2023_PySR}), scikit-learn (\cite{scikit-learn}), and SciPy (\cite{scipy}).

FM thanks the Institut Philippe Meyer for financial support. LCP thanks the Natural Sciences and Engineering Research Council of Canada for funding.

\end{acknowledgments}

\begin{contribution}

F.Masson analyzed the data and wrote the draft. L.C.Parker supervised the research, edited and reviewed the paper.

\end{contribution}

\appendix

\section{RMSE in Projected Phase Space at Different Redshifts} \label{app:rmse(PPS)(z)}

To complement the analysis of RMSE presented in Section~\ref{subsubsec:rmse_z0}, Fig.~\ref{fig:rmse(z)PPS} shows the spatial distribution of RMSE in PPS at six redshifts, and Fig.~\ref{fig:fracError(z)PPS} shows the spatial distribution of the fractional error $\frac{\rm RMSE}{T_{\rm inf}}$ in PPS at the same redshifts.

\begin{figure}[htbp]
    \centering
    \begin{subfigure}{0.31\textwidth}
        \includegraphics[width=\linewidth]{PPS_RMSE_snap99_corrected.pdf}
        \captionsetup{labelformat=empty}
        \caption{$z=0$}
    \end{subfigure}
    \hfill
    \begin{subfigure}{0.31\textwidth}
        \includegraphics[width=\linewidth]{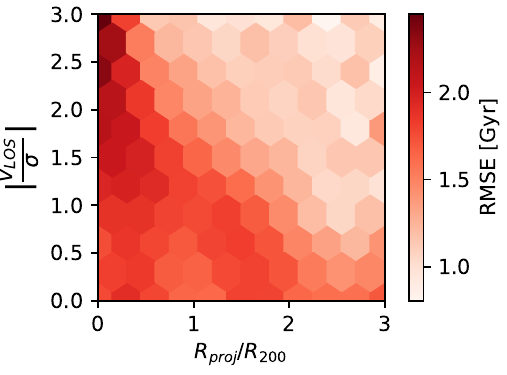}
        \captionsetup{labelformat=empty}
        \caption{$z=0.2$}
    \end{subfigure}
    \hfill
    \begin{subfigure}{0.31\textwidth}
        \includegraphics[width=\linewidth]{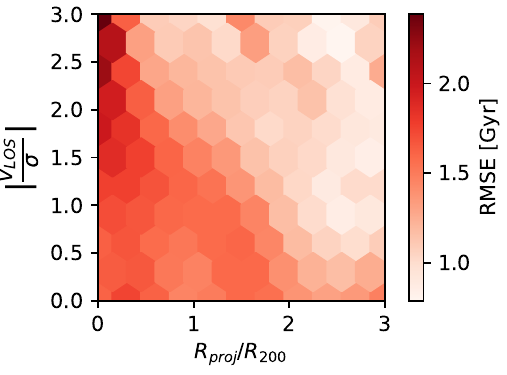}
        \captionsetup{labelformat=empty}
        \caption{$z=0.3$}
    \end{subfigure}
    
    \begin{subfigure}{0.31\textwidth}
        \includegraphics[width=\linewidth]{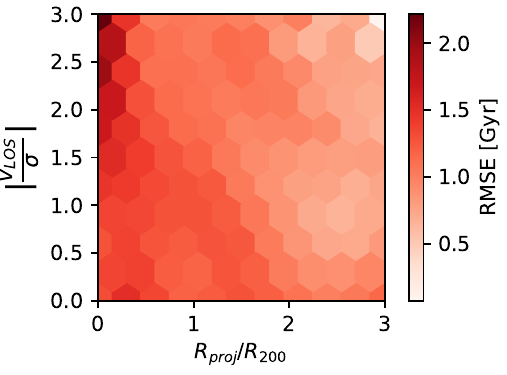}
        \captionsetup{labelformat=empty}
        \caption{$z=0.5$}
    \end{subfigure}
    \hfill
    \begin{subfigure}{0.31\textwidth}
        \includegraphics[width=\linewidth]{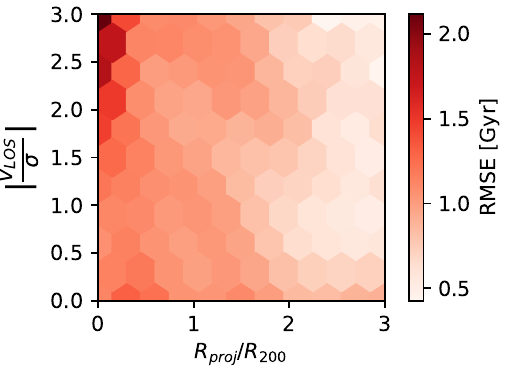}
        \captionsetup{labelformat=empty}
        \caption{$z=0.7$}
    \end{subfigure}
    \hfill
    \begin{subfigure}{0.31\textwidth}
        \includegraphics[width=\linewidth]{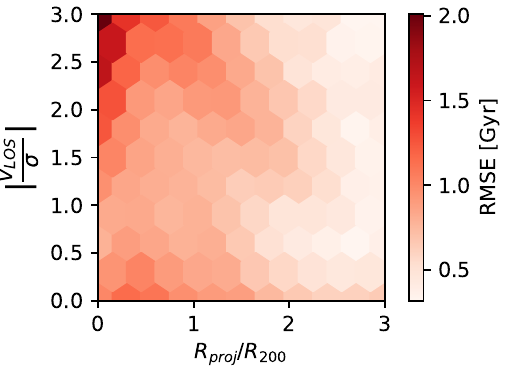}
        \captionsetup{labelformat=empty}
        \caption{$z=1$}
    \end{subfigure}
    \caption{RMSE in projected phase space at six different redshifts.}
    \label{fig:rmse(z)PPS}
\end{figure}

\begin{figure}[htbp]
    \centering
    \begin{subfigure}{0.31\textwidth}
        \includegraphics[width=\linewidth]{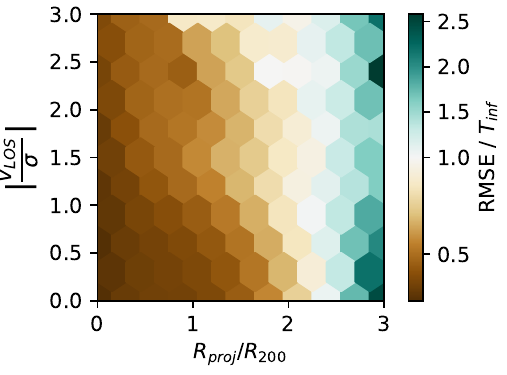}
        \captionsetup{labelformat=empty}
        \caption{$z=0$}
    \end{subfigure}
    \hfill
    \begin{subfigure}{0.31\textwidth}
        \includegraphics[width=\linewidth]{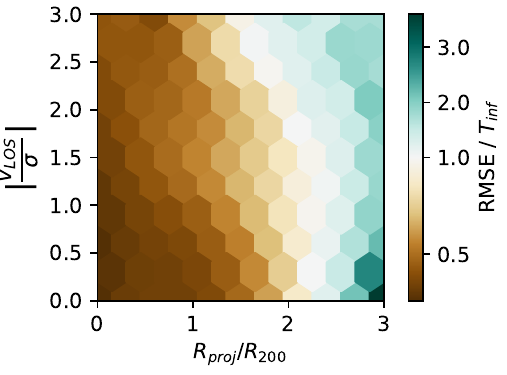}
        \captionsetup{labelformat=empty}
        \caption{$z=0.2$}
    \end{subfigure}
    \hfill
    \begin{subfigure}{0.31\textwidth}
        \includegraphics[width=\linewidth]{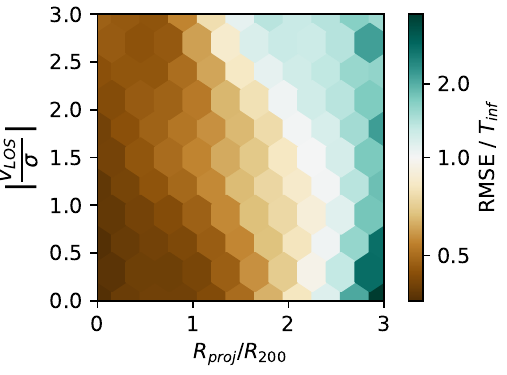}
        \captionsetup{labelformat=empty}
        \caption{$z=0.3$}
    \end{subfigure}
    
    \begin{subfigure}{0.31\textwidth}
        \includegraphics[width=\linewidth]{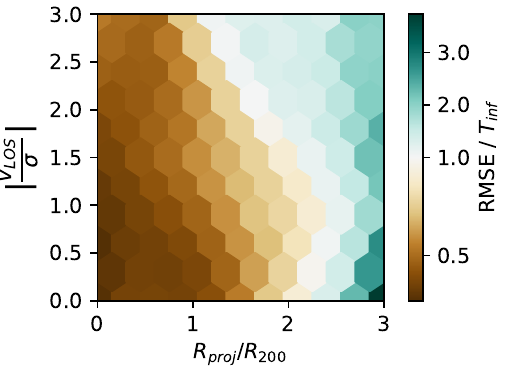}
        \captionsetup{labelformat=empty}
        \caption{$z=0.5$}
    \end{subfigure}
    \hfill
    \begin{subfigure}{0.31\textwidth}
        \includegraphics[width=\linewidth]{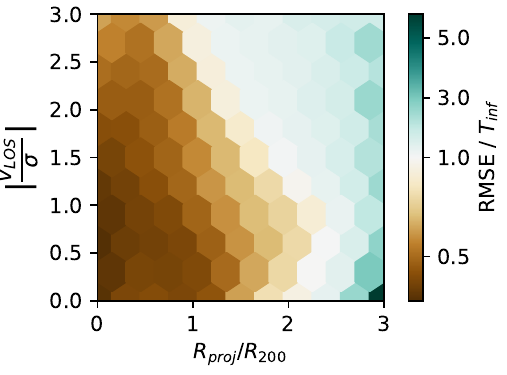}
        \captionsetup{labelformat=empty}
        \caption{$z=0.7$}
    \end{subfigure}
    \hfill
    \begin{subfigure}{0.31\textwidth}
        \includegraphics[width=\linewidth]{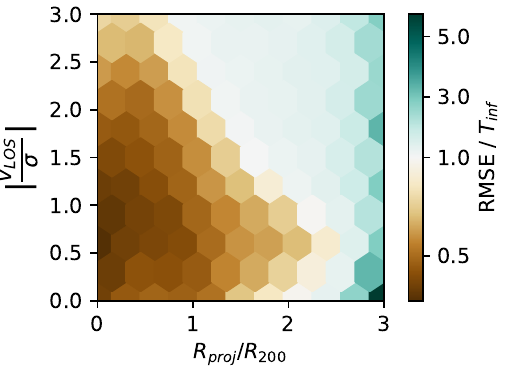}
        \captionsetup{labelformat=empty}
        \caption{$z=1$}
    \end{subfigure}
    \caption{Fractional error $\frac{\rm RMSE}{T_{\rm inf}}$ in projected phase space at six different redshifts. The colorbars are linear for $\frac{\rm RMSE}{T_{\rm inf}} \in [0,1]$ and for $\frac{\rm RMSE}{T_{\rm inf}} > 1$, but the slopes of these two portions are different for better clarity.}
    \label{fig:fracError(z)PPS}
\end{figure}

\section{Calibration without the $M_*$-dependence} 
\label{app:MsDependence}

Here we provide the mean infall times and RMSE without an explicit dependence on galaxy stellar mass as described in Section \ref{sec:finalformula}.

\begin{deluxetable*}{lcccccccc}
\tablecaption{Mean infall times in the \cite{Pasquali_2019} (P19) zones and RMSE of our calibration as a function of redshift without explicit dependence on stellar mass (see Eq.~(\ref{eq:Tinf(z)_meanMs})).\label{tab:rmseMsFixed}}

\tablehead{
\colhead{Zone\tablenotemark{a}} &
\colhead{$\overline{T}_{\mathrm{inf,eq},z=0}$\tablenotemark{b}} &
\colhead{$\overline{T}_{\mathrm{inf,eq}, z=0,M_*}$\tablenotemark{c}} &
\multicolumn{6}{c}{$\sigma_{T_{\mathrm{inf}, M_*}}$ [Gyr]\tablenotemark{d}} \\
\colhead{} &
\colhead{[Gyr]} & \colhead{[Gyr]} & \colhead{$z=0.0$} & \colhead{$z=0.2$} &
\colhead{$z=0.3$} & \colhead{$z=0.5$} & \colhead{$z=0.7$} & \colhead{$z=1.0$}
}
\startdata
1 & 8.39 & 8.13 & 2.33 & 1.82 & 1.66 & 1.35 & 1.11 & 0.90 \\
2 & 7.66 & 7.37 & 2.47 & 1.95 & 1.76 & 1.46 & 1.24 & 0.98 \\
3 & 6.98 & 6.70 & 2.47 & 1.93 & 1.75 & 1.45 & 1.26 & 1.02 \\
4 & 6.20 & 5.93 & 2.42 & 1.86 & 1.65 & 1.36 & 1.16 & 0.94 \\
5 & 5.70 & 5.43 & 2.49 & 1.90 & 1.70 & 1.40 & 1.20 & 0.92 \\
6 & 5.38 & 5.11 & 2.49 & 1.91 & 1.70 & 1.38 & 1.21 & 0.94 \\
7 & 5.12 & 4.86 & 2.39 & 1.82 & 1.62 & 1.32 & 1.15 & 0.91 \\
8 & 4.69 & 4.42 & 2.08 & 1.59 & 1.43 & 1.27 & 1.23 & 1.09 \\
\enddata
\tablenotetext{a}{Zone number from P19.}
\tablenotetext{b}{Mean infall time computed with Eq.~\ref{eq:Tinf} at $z=0$.}
\tablenotetext{c}{Mean infall time computed with Eq.~\ref{eq:Tinf(z)_meanMs} at $z=0$.}
\tablenotetext{d}{RMSE at each redshift in each zone with $T_{\rm inf}$ computed with Eq.~(\ref{eq:Tinf(z)_meanMs}).}
\end{deluxetable*}

\section{Gaussian Fit Parameters}
\label{app:gaussianParameters}

The black curves in Fig.~\ref{fig:gridHist} can be expressed as: 

\[ 
f(T_{\rm inf}) = w_1e^{-\frac{(T_{\rm inf}-\mu_1)^2}{\sigma_1^2}} + w_2e^{-\frac{(T_{\rm inf}-\mu_2)^2}{\sigma_2^2}}
\]

with the parameters for each cell provided in Table~\ref{tab:grid66gaussians}.

\begin{deluxetable}{cccccccccc}
\tablecaption{Gaussian fitting parameters for the distributions presented in Figure~\ref{fig:gridHist}.\label{tab:grid66gaussians}}
\tablehead{
\colhead{$r_{\rm min}$} &
\colhead{$r_{\rm max}$} &
\colhead{$v_{\rm min}$} &
\colhead{$v_{\rm max}$} &
\colhead{$\mu_1$} &
\colhead{$\sigma_1$} &
\colhead{$w_1$} &
\colhead{$\mu_2$} &
\colhead{$\sigma_2$} &
\colhead{$w_2$} 
}
\startdata
0.00 & 0.50 & 0.00 & 0.50 & 5.34 & 1.65 & 0.38 & 9.25 & 1.33 & 0.62 \\
0.00 & 0.50 & 0.50 & 1.00 & 5.06 & 1.64 & 0.37 & 9.15 & 1.36 & 0.63 \\
0.00 & 0.50 & 1.00 & 1.50 & 4.50 & 1.65 & 0.37 & 8.88 & 1.45 & 0.63 \\
0.00 & 0.50 & 1.50 & 2.00 & 4.18 & 1.62 & 0.42 & 8.79 & 1.42 & 0.58 \\
0.00 & 0.50 & 2.00 & 2.50 & 3.80 & 1.33 & 0.46 & 8.55 & 1.47 & 0.54 \\
0.00 & 0.50 & 2.50 & 3.00 & 3.72 & 1.21 & 0.51 & 8.44 & 1.47 & 0.49 \\
0.50 & 1.00 & 0.00 & 0.50 & 4.79 & 1.49 & 0.53 & 8.32 & 1.31 & 0.47 \\
0.50 & 1.00 & 0.50 & 1.00 & 4.32 & 1.50 & 0.51 & 7.99 & 1.40 & 0.49 \\
0.50 & 1.00 & 1.00 & 1.50 & 3.60 & 1.54 & 0.56 & 7.57 & 1.55 & 0.44 \\
0.50 & 1.00 & 1.50 & 2.00 & 3.31 & 1.38 & 0.68 & 7.22 & 1.59 & 0.32 \\
0.50 & 1.00 & 2.00 & 2.50 & 3.30 & 1.27 & 0.80 & 6.72 & 1.68 & 0.20 \\
0.50 & 1.00 & 2.50 & 3.00 & 3.16 & 1.13 & 0.58 & 3.97 & 2.05 & 0.42 \\
1.00 & 1.50 & 0.00 & 0.50 & 3.25 & 1.40 & 0.32 & 6.51 & 1.52 & 0.68 \\
1.00 & 1.50 & 0.50 & 1.00 & 2.67 & 1.26 & 0.39 & 6.30 & 1.53 & 0.61 \\
1.00 & 1.50 & 1.00 & 1.50 & 2.21 & 1.14 & 0.56 & 5.89 & 1.56 & 0.44 \\
1.00 & 1.50 & 1.50 & 2.00 & 2.13 & 1.09 & 0.65 & 5.06 & 1.62 & 0.35 \\
1.00 & 1.50 & 2.00 & 2.50 & 1.83 & 0.96 & 0.57 & 3.87 & 1.50 & 0.43 \\
1.00 & 1.50 & 2.50 & 3.00 & 1.33 & 0.77 & 0.61 & 3.08 & 1.42 & 0.39 \\
1.50 & 2.00 & 0.00 & 0.50 & 1.94 & 0.88 & 0.43 & 6.23 & 1.27 & 0.57 \\
1.50 & 2.00 & 0.50 & 1.00 & 1.81 & 0.95 & 0.59 & 6.16 & 1.31 & 0.41 \\
1.50 & 2.00 & 1.00 & 1.50 & 1.66 & 0.97 & 0.78 & 5.67 & 1.53 & 0.22 \\
1.50 & 2.00 & 1.50 & 2.00 & 1.47 & 0.86 & 0.73 & 3.90 & 1.75 & 0.27 \\
1.50 & 2.00 & 2.00 & 2.50 & 1.17 & 0.80 & 0.73 & 3.09 & 1.65 & 0.27 \\
1.50 & 2.00 & 2.50 & 3.00 & 0.90 & 0.64 & 0.73 & 2.61 & 1.54 & 0.27 \\
2.00 & 2.50 & 0.00 & 0.50 & 1.26 & 0.73 & 0.71 & 6.06 & 1.46 & 0.29 \\
2.00 & 2.50 & 0.50 & 1.00 & 1.09 & 0.68 & 0.75 & 5.07 & 1.92 & 0.25 \\
2.00 & 2.50 & 1.00 & 1.50 & 0.99 & 0.68 & 0.77 & 3.94 & 1.91 & 0.23 \\
2.00 & 2.50 & 1.50 & 2.00 & 0.79 & 0.55 & 0.70 & 2.87 & 1.56 & 0.30 \\
2.00 & 2.50 & 2.00 & 2.50 & 0.73 & 0.54 & 0.75 & 2.78 & 1.56 & 0.25 \\
2.00 & 2.50 & 2.50 & 3.00 & 0.64 & 0.41 & 0.75 & 2.32 & 1.43 & 0.25 \\
2.50 & 3.00 & 0.00 & 0.50 & 0.44 & 0.38 & 0.77 & 4.22 & 2.43 & 0.23 \\
2.50 & 3.00 & 0.50 & 1.00 & 0.42 & 0.38 & 0.76 & 3.57 & 2.24 & 0.24 \\
2.50 & 3.00 & 1.00 & 1.50 & 0.39 & 0.36 & 0.70 & 2.91 & 1.97 & 0.30 \\
2.50 & 3.00 & 1.50 & 2.00 & 0.32 & 0.30 & 0.68 & 2.42 & 1.72 & 0.32 \\
2.50 & 3.00 & 2.00 & 2.50 & 0.30 & 0.28 & 0.72 & 2.53 & 1.90 & 0.28 \\
2.50 & 3.00 & 2.50 & 3.00 & 0.25 & 0.25 & 0.69 & 2.40 & 1.72 & 0.31 \\
\enddata
\tablenotetext{}{The first four columns define the cell boundaries. The remaining columns list the parameters used to model the infall‐time distributions in these cells as the sum of two Gaussians.}
\end{deluxetable}

\bibliography{biblioInfallTime}{}
\bibliographystyle{aasjournalv7}

\end{document}